\documentclass[draftcls,12pt,onecolumn]{IEEEtran}
\PassOptionsToPackage{bookmarks={false}}{hyperref}
\usepackage{amssymb}
\usepackage{cite}
\usepackage{url}
\usepackage{xcolor}
\usepackage{amsmath}
\usepackage{graphicx}
\usepackage{derivative}
\usepackage{algorithm}
\usepackage{algpseudocode}
\usepackage{mathtools}
\usepackage{balance}
\usepackage{commath}
\usepackage{epstopdf}
\usepackage{blkarray} %
\usepackage{adjustbox}
\usepackage{bm}

\begin{document}

	\title{Movable Antenna for Integrated Sensing and
		Communication in Air–Sea-Ground Networks}
	\author{Ahmed A. Al-habob,~\IEEEmembership{Senior Member,~IEEE,} 
	Octavia A. Dobre,~\IEEEmembership{Fellow,~IEEE,}  and 
	Yindi Jing
}

	\maketitle
	
\IEEEpeerreviewmaketitle
\begin{abstract}
	  Integrated sensing and communication (ISAC) is a new paradigm for efficiently combining sensing and communication functionalities by leveraging shared hardware and radio resources.  Despite its promise, ISAC yields conflicting beamforming goals and competition over the same resources. Movable antennas enable effective exploitation of spatial degrees of freedom through dynamic position/orientation control, thereby enhancing the performance of ISAC systems.
	This paper proposes a movable antenna framework for ISAC in air–sea-ground  networks. A multi-objective optimization problem is formulated with the objectives of maximizing the communication  rate  of a set of aerial, sea, and ground devices and the   sensing rate of a set of targets. The location and orientation of the antenna sub-arrays, as well as the transmit/receive beamforming, are optimized under practical constraints on the movable antennas' location and orientation.
	 A  solution  is developed based on a $K$-means clustering approach to optimize the sub-arrays' orientation and a particle swarm optimization to place the sub-arrays in optimized locations. The transmit  and receive beamforming   are designed   using a successive convex approximation and a  generalized eigenvector  method, respectively.  Simulation results illustrate that the developed  movable antenna framework    improves the ISAC objective and provides a remarkable trade-off between the communication data rate and the targets' sensing rate when compared with the conventional stationary antenna array scenario. 
\end{abstract}
 \section{Introduction}
 Communication networks have been evolving toward the sixth generation networks, which are associated with a   rapid development of non-terrestrial network   technologies  to provide global connectivity to areas and regions   that cannot be covered by existing terrestrial networks \cite{10560514}.  For instance, maritime exploration and activities such as offshore aquaculture, marine tourism, and oceanic mineral extraction are essential to the blue economy and the sustainable use of ocean resources. 
 These activities impose the need for reliable communication with surface stations such as  vessels, buoys, and offshore platforms as well as energy-efficient data gathering   \cite{940914877}. Another component of the non-terrestrial networks  is
airborne platforms, which  exploits line-of-sight (LoS) as well as  non line-of-sight (NLoS) links.  Unmanned aerial vehicles (UAVs) are examples of the airborne platforms that provide a flexible and cost-effective option for communication coverage improvement and data   aggregation \cite{921706688}.  It is worth mentioning that the integration of the ground, sea, and aerial networks is a cost-efficient attempt to improve the data rate on the ocean and in  remote rural areas, in which the achieved data rate    is way below
that of the fifth-generation cellular networks.  Such network integration increases the number of devices and their heterogeneity, and the devices' locations are diverse, which is an additional challenge for the base stations.   Embedding more antennas to the base stations  
is considered to improve  the communication channel  which has driven the development of  multiple-input and multiple-output (MIMO),  massive MIMO (mMIMO), and extreme MIMO (X-MIMO)  technologies \cite{11503880}.
However, these technologies   have fixed
antenna orientations and/or positions once deployed,  which restricts their performance in the spatially diverse scenarios. 
 Fixed antenna systems exhibit static radiation patterns, which require a large number of antenna elements to track rapid channel variations across different multipath environments and high-mobility scenarios.  Furthermore, the performance of high-frequency communication and target-sensing systems is susceptible to severe path loss and directional misalignment, which impose additional challenges on the design of fixed antenna systems.

Movable antenna emerges as a promising technique to improve the communication channel between the base station and the receiver  \cite{10945745,jiang2025statistical}.
Unlike the conventional fixed antenna placement paradigm,  movable antennas can   change  orientation and location   to improve the communication channel with the users \cite{jiang2025statistical}. Therefore,  the movable antenna paradigm can provide better performance
using the same  number of antenna elements.
 Furthermore, the movable antenna paradigm coexists with most of the existing beamforming and antenna   techniques \cite{shao2025tutorial}. 
Movable antennas are part of a recent research effort, namely reconfigurable antenna systems that include fluid antennas (FAs), reconfigurable holographic antennas (RHAs), pinching antennas (PAs), and six-dimensional movable antennas (6DMAs) \cite{11433651}.  FAs systems utilize liquid metals or conductive electrolytes as radiating elements. The electrical properties of the antenna elements can be adjusted by controlling the fluid volume or position.  This flexibility makes FAs suitable for adaptable and compact applications such as
wearable devices and small-form-factor sensors. It is worth mentioning that FAs performance depends on the fluid
stability under different environmental conditions, which requires precise control mechanisms to maintain consistent reconfiguration \cite{9264694}.
Based on holography
principles, the radiation pattern of an RHA element can be adjusted by changing the phase and amplitude of the incident
electromagnetic waves  using metasurfaces or
diffractive optical elements.   
This enables RHAs to be suitable for high-capacity networks, where precise radiation pattern control is crucial for efficient
spectrum utilization. On the other hand, RHAs design and fabrication are complex processes,  involving advanced techniques and a high-precision design \cite{10163760}. 
PAs systems  involve mechanical deformation, such as pinching or squeezing, to alter electrical properties and control the radiation pattern. 
While providing remarkable performance with a low-complexity reconfiguration, the mechanical nature and reconfiguration range of PAs may be adjusted along a limited pre-installed waveguide. \cite{10945421}.
The antenna elements (or antenna subarrays) in 6DMAs systems change position and/or orientation through mechanical means such as motors and actuators \cite{11142311}. This adaptability is beneficial in scenarios involving changes in the devices' position   by adding six-dimensional  positions and rotations (3D rotations and 3D
positions)  to the antenna placement at the base station and/or the receiver \cite{jiang2025statistical}. Furthermore, it enables the base station to perform multiple tasks such as communication and sensing moving targets \cite{10945745}.

Integrated sensing and communication (ISAC) is another emerging technology, which allows the   sharing of the signal processing modules and hardware platforms for radar sensing  and communication  functions \cite{9737357}. 
Through appropriate
   resource utilization and strategic module collaboration,  ISAC enhances   localization,   communication, and covert communication performance \cite{10901459}. 
    Conventional stationary antennas   are  widely
  employed and configured along with waveform or precoding matrix   programming   to improve the communication channel capacity and
  sensing accuracy.  Numerous research works have been devoted to explore antenna
  placement optimization techniques to improve  spatial
  diversity of  wireless channels. 
  Antenna selection is one of these attempts, in which   a subset of
  antennas are dynamically selected and activated to enhance the communication and sensing.
  Array synthesis is another attempt to add degrees of freedom, enhance beam pattern indicators, minimize sidelobe power, and maintain good array sparsity in ISAC scenarios.      
However,	ISAC poses challenges, including conflicting beamforming goals, additional constraints on resource allocation, and limitations in hardware and propagation conditions.   Furthermore, practical ISAC deployments are sensitive to the channel impairments, imperfect alignment, and user mobility, which necessitate innovative architectures beyond classical fixed-antenna and beamforming approaches \cite{9705498}. By controlling the position and orientation of movable antenna systems, the channel matrix can be reconfigured to enhance the channel
	quality and improve communication and sensing. This adaptability enables movable antenna systems to maximize spatial diversity in multipath environments and provides accurate beam alignment for both communication and sensing. This is beneficial for integrated networks that involve a large number of heterogeneous devices, with high spatial diversity.

\subsection{Related Work}

In the existing literature, research on air–sea-ground  networks was primarily
dedicated to  network architecture design \cite{9453860,9869801,10269648} and    resource management and allocation  \cite{10083276,10454605}. In \cite{9453860}, a hybrid
space-air-ground network was considered  for maritime coverage enhancement.  Communication link assignment and rate adaptation techniques were developed  to minimize the  consumed energy  under link quality-of-service requirements. 
In \cite{9869801}, a  space-air-ground-sea integrated network   architecture was studied to improve the     signal coverage and provide high-quality   service for terrestrial and maritime areas.   A   satellite network structure  to serve aviation users   and marine users was considered in \cite{10269648}. Heat maps were created to generate a benchmark observation point model to provide   maximum coverage with minimum number of  satellites. 
 In \cite{10083276}, an edge computing  resource slicing problem in  space-air-ground-sea  edge computing   network was studied. The network consists of multiple moving air-ground-sea nodes with unstable communication links and two possible states: reliable, which represents an error-free transmission, and unreliable, which indicates that random errors are propagated/generated by the node. The  nodes’ states were estimated using a  message passing graph neural network. 
In \cite{10454605}, a distributed uplink scheduling framework was developed for space–air–ground–sea integrated networks. In this framework, the access tier, spectrum slice, and transmit power were assigned to maritime nodes under interference, energy, and spectrum deadline constraints. A game-theoretic   multi-agent reinforcement learning algorithm was designed to perform the resource scheduling.

Beamforming design for ISAC scenarios was primarily studied   with fixed-antenna systems \cite{10027173, 10497104, 11296885, 10902607}. In \cite{10027173}, a beamforming design approach was studied for a MIMO dual-function communication-radar scenario with direction mismatch and communication channel-state information error. The radar sensing performance was evaluated in a region-of-interest, while the communication was assumed to be subject to  unknown errors. The beamforming approach aimed to optimize radar sensing performance by balancing the radiated energy  toward the region-of-interest with  communication user requirements.  A   semidefinite relaxation technique was adopted as a solution algorithm. In \cite{10497104}, a transceiver design for a covert ISAC  system was designed. A worst-case and outage-constrained optimization problem was formulated to design transceiver beamforming and radar waveforms with the objective of balancing multi-target sensing while maintaining communication covertness requirements.   An alternating optimization solution was proposed to decouple  the transceiver beamforming into transmitter and receiver feasibility-checking subproblems. In \cite{11296885}, a cell-free mMIMO  ISAC scenario was considered, in which a set of access points performs ISAC tasks by providing communication services to users and sensing a set of targets. Detection/tracking approaches based on a likelihood-ratio-rest technique were developed to handle unknown target responses and interference. A power control mechanism was also developed to balance the power allocation for communication and sensing objectives. In \cite{10902607}, a predictive beamforming framework for a secure  ISAC scenario was proposed, in which a set of UAV eavesdroppers attempts to intercept the communication from a base station to a set of users. The eavesdroppers' locations are variable and unknown, and must be estimated from the echo signal. A coarse estimation with refinements was developed to estimate  the    location of   each eavesdropper, and a deep network was trained to predict their channels. Based on the predicted eavesdroppers’ channels, two secure beamforming approaches were designed to enhance the sum secrecy rate for the users.

\begin{table*}[h!] 
	\caption{Description of Main Notations.}
	\begin{center}
		\begin{adjustbox}{width=.99\textwidth}
			\begin{tabular}{|c|c||c|c|}
				\hline
				{Notation} &  {Description} & {Notation} &  {Description}\\ \hline 
				$  T $	&\multicolumn{1}{l||}{Number of  targets in the targets set $  \mathcal{T}  $}  & $ U $ & \multicolumn{1}{l|}{Number of  UAVs in the UAVs set $\mathcal{U}$}           \\ \hline   
				$  S $	&\multicolumn{1}{l||}{Number of  sea surface stations in the set $  \mathcal{S}  $}  & $ G $ & \multicolumn{1}{l|}{Number of  ground users in the  set $\mathcal{G}$}           \\ \hline   
				$  D = U+S+G$	&\multicolumn{1}{l||}{Total number of  devices in the set $\mathcal{D}\triangleq\mathcal{U}\cup \mathcal{S}\cup   \mathcal{G}$}  & $ \bar{\bm{\psi}}_i = [\bar{x}_i, \bar{y}_i, \bar{z}_i]^T $ & \multicolumn{1}{l|}{The  coordinates of the $i$-th device }           \\ \hline   
				$  K$	&\multicolumn{1}{l||}{Number of  antenna sub-arrays}  & $ N_k $ & \multicolumn{1}{l|}{Number of antenna elements in the $k$-th sub-array }           \\ \hline   
				$  N$	&\multicolumn{1}{l||}{Total number of  antenna elements  $N=\sum_{k=1}^{K}N_k$}  & $ \bm{\psi}_k = [x_k, y_k, z_k]^T $ & \multicolumn{1}{l|}{The coordinates of the $k$-th    sub-array}           \\ \hline   
				$  \bm{\varphi}_k = [\alpha_k, \beta_k, \gamma_k]^T$	&\multicolumn{1}{l||}{Rotation angles around  the $x$-axis, $y$-axis and $z$-axis}  & $\mathcal{C}$ & \multicolumn{1}{l|}{3D space around the base station }           \\ \hline   
				$\textbf{R}\left(\bm{\varphi}_k\right)$	&\multicolumn{1}{l||}{The rotation matrix }  & $\tilde{\textbf{r}}_n/\textbf{r}_{k,n}(\bm{\psi}_k,\bm{\varphi}_k)$ & \multicolumn{1}{l|}{Position of the $n$-th antenna in the $k$-th sub-array in the local/global coordinates}           \\ \hline   
				$\textbf{a}(\!\bm{\psi}_k,\!\bm{\varphi}_k\!)$	&\multicolumn{1}{l||}{The steering vector of the $k$-th sub-array}  & $\tilde{\textbf{n}}_k/{\textbf{n}}_k\left(\bm{\varphi}_k\right)  $ & \multicolumn{1}{l|}{Outward normal vector of the $k$-th sub-array  in the local/global coordinates}           \\ \hline   
				$\textbf{f}_k$ 	&\multicolumn{1}{l||}{The	pointing vector along $(\theta_k, \phi_k)$}  & $\chi_{k,i}$ & \multicolumn{1}{l|}{The directive   gain of the $k$-th antenna sub-array towards the $i$-th device }           \\ \hline   
				$\mathcal{A}$	&\multicolumn{1}{l||}{Maximum antenna gain}  &  $\varpi_1/\varpi_2$ & \multicolumn{1}{l|}{Front-back ratio/sidelobe level}           \\ \hline   
				$\mathcal{A}_{h_{k,i}}/\mathcal{A}_{v_{k,i}}$	&\multicolumn{1}{l||}{Horizontal/vertical radiation pattern}  & $(\tilde{\theta}_k, \tilde{\phi}_k)$ & \multicolumn{1}{l|}{Direction of departure in the local coordinates}           \\ \hline 
				$(\tilde{\theta}_{k,i}, \tilde{\phi}_{k,i})$	&\multicolumn{1}{l||}{Direction of $i$-th device/target in the local coordinates}  & $\phi_{3\mbox{\scriptsize dB}}/\theta_{3\mbox{\scriptsize dB}} $ & \multicolumn{1}{l|}{The horizontal/vertical  $3$-dB beamwidth}           \\ \hline   
				$\bm{\Psi} \triangleq [\bm{\psi}_1^T, \cdots, \bm{\psi}_K^T]^T$	&\multicolumn{1}{l||}{Sub-arrays placement decision variable}  & $\bm{\Upsilon} \triangleq [\bm{\varphi}_1^T, \cdots, \bm{\varphi}_K^T]^T$ & \multicolumn{1}{l|}{Sub-arrays rotation decision variable}           \\ \hline    
				$\textbf{h}_i(\bm{\Psi},\bm{\Upsilon})$	&\multicolumn{1}{l||}{Communication   channel  between  the base station and the $i$-th device}  & $\Lambda_{k,i}(\bm{\psi}_k)$ & \multicolumn{1}{l|}{Channel gain between the  $k$-th sub-array and $i$-th device}           \\ \hline     
				$\delta_{k,i}=\| \bm{\psi}_k- \bar{\bm{\psi}}_i\|_2 $	&\multicolumn{1}{l||}{Distance from the $k$-th sub-array to the $i$-th device}  & $\vartheta_u$ & \multicolumn{1}{l|}{Path loss at a distance of $1$ meter}           \\ \hline       
				$\vartheta_0 $	&\multicolumn{1}{l||}{NLoS attenuation factor}  & $\ell_u$ & \multicolumn{1}{l|}{Path loss exponent}           \\ \hline   
				$ {\Pr}_{k,i}\left(\mbox{\small LoS}\right)/F$	&\multicolumn{1}{l||}{Probability of a LoS connection/Rician factor}  & $h_i/h_k$ & \multicolumn{1}{l|}{Height of the $i$-th surface station/$k$-th sub-array}           \\ \hline     
				$\omega$	&\multicolumn{1}{l||}{Relative weight
					to maximizing the communication  or  sensing sum  rate}  & $ \tilde{{h}}^{\mbox{\scriptsize NLoS}}_{k,i} \sim \mathcal{N}(0,1)  $ & \multicolumn{1}{l|}{NLoS   scattering component}           \\ \hline   
				$\mathbf{w}_i/\mathbf{\bar{w}}_t$	&\multicolumn{1}{l||}{Beamformer for the $i$-th device/$t$-th target}  & $\Gamma_i(\bm{\Psi},\bm{\Upsilon},\mathbf{w})/\sigma^2_i$ & \multicolumn{1}{l|}{SINR/power of the additive noise   at the $i$-th device }           \\ \hline         
				$\textbf{H}_t(\bm{\Psi},\bm{\Upsilon})$	&\multicolumn{1}{l||}{Sensing channel of the $t$-th target}  & $\Lambda_{k,t}(\bm{\psi}_k)$ & \multicolumn{1}{l|}{The reflected complex amplitude of the $t$-th target at the $k$-th sub-array}           \\ \hline   
				$ \mathbf{{q}}_t$	&\multicolumn{1}{l||}{Receive beamformer   for the $t$-th target}  & $\bar{\Gamma}_t(\bm{\Psi},\bm{\Upsilon},\mathbf{w},\mathbf{q})/\sigma^2_r$ & \multicolumn{1}{l|}{The SINR/power of the additive noise   of the $t$-th target }           \\ \hline       
				$ \bar{\mathcal{R}}(\bm{\Psi},\bm{\Upsilon},\mathbf{w},\mathbf{q})$	&\multicolumn{1}{l||}{Sum sensing rate}  & $\Omega$ & \multicolumn{1}{l|}{Number of particles in the  particles set $\bm{\varOmega}$}           \\ \hline                                                                                                                                                                                                                                                                                                                                                                                                                                                                                                                                                                                                                                                                                                                                                                                                                                                                                                                                                                                                                                                                                                                                                                                                                                                                       
			\end{tabular}
		\end{adjustbox}
	\end{center}
	\label{mysymbols}
\end{table*}

Recently, research works have been dedicated to leveraging more degrees-of-freedom   in the design of high-performance communication and more accurate sensing using  movable antenna systems \cite{10643473, 10944486, 11075964, 10839251,11283108,11493574,11249714,11177504}. 
In \cite{10643473}, a  wireless sensing system equipped with movable antennas was explored, in which the positions of antenna elements were   dynamically adjusted in  one-dimensional (1D) as well as two-dimensional
(2D) movable antenna arrays to improve the sensing performance. In \cite{10944486}, a 2D movable antenna framework was developed with the objective of   maximizing   the sensing signal-to-interference-plus-noise ratio (SINR) while maintaining   the SINR  for the communication devices above a given threshold. In \cite{11075964}, 
a near-field ISAC system was considered, in which multiple 2D transmit and receive movable antennas are placed at the base station   to transmit data to   a set of devices and to sense a set of targets. In \cite{10839251}, a 1D movable antenna ISAC framework was developed to enable communication with a set of users and sensing a target by adjusting the antenna coefficients and the antenna placement. In \cite{11283108}, a resource allocation problem for a 2D movable antenna-enabled ISAC scenario was studied. A two-timescale optimization framework was designed to adjust the discretized   antenna placement, the beamforming vectors, and the snapshot duration. The considered objective was to minimize the transmit power at the base station while maintaining minimum communication and sensing   requirements. 
In \cite{11493574}, a 1D movable antenna-enabled intelligent reflecting surface (IRS)-aided ISAC scenario was investigated. In this scenario,  a
base station with a linear movable antenna system communicates with a group of devices and estimates the direction-of-arrival of
a target. An optimization problem was formulated with  maximizing the communication sum rate of the devices as an objective by adjusting the antenna locations, beamforming, and the  IRS parameters, subject to constraints on the sensing requirements. A  Riemannian manifold-based method was designed to solve the optimization problem.
In \cite{11249714}, a full-duplex 2D movable antenna  ISAC  system was considered, in which a base station communicates with a set of uplink and downlink users and senses a set of targets.  The transmit power consumption was minimized by optimizing the
 discrete position of the movable antenna elements, beamforming vectors, and sensing signal covariance matrix, subject to constraints on the communication and sensing requirements. 
In \cite{11177504}, a secure communication problem in an IRS-assisted ISAC system was considered, in which a base station with a 1D movable antenna system transmits information to a legitimate receiver in the presence of an eavesdropping target with the help of an IRS array. Maximizing the secrecy rate was considered as an objective, subject to constraints on the available power and radar beam requirements.

Motivated by the above discussions,   this paper  considers  a movable antenna-based framework   to optimize the transmit/receive beamforming, as well as the location and the orientation of a set of antenna sub-arrays at the base station. The base station senses a set of targets and communicates with a set of UAVs, sea surface stations, and ground users.  
The main contributions of this paper can be summarized as follows:

\begin{itemize}
	\item[$\bullet$] A movable antenna-based framework for ISAC scenario in air–sea-ground  networks is developed.
	\item A multi-objective optimization problem is formulated to maximize both the communication data rate and the sensing rate with practical constrains on the sub-arrays' placement and orientation.
	\item[$\bullet$]  A  solution approach is developed to solve the optimization problem, in which the sub-arrays' orientation is obtained based on a $K$-means clustering technique, the  sub-arrays' locations are determined through a particle swarm optimization (PSO), and  the receive and  transmit   beamforming are obtained using a  generalized eigenvector and a successive convex approximation approaches, respectively.
\end{itemize}

The remainder of this paper is structured as follows.
Section \ref{Sys}  introduces the considered system model. Section \ref{ProbFormulation} discusses the multi-objective optimization  problem formulation.  Section \ref{Sol2} presents the proposed  solution and its computational complexity.     
Section \ref{SimSec} illustrates and discusses the   simulation
results, while Section \ref{CO} concludes the paper.

\textit{Notation:} Upper case boldface letters and lower case boldface letters represent   matrices and vectors, respectively. 	$[\cdot]^T$  and $[\cdot]^H$    denote the transpose  and the Hermitian       operators, respectively. $\mathcal{B} \cup \mathcal{E}$ is the union of   set  $\mathcal{B}$ and set $\mathcal{E}$ while  $\mathcal{B} \setminus \mathcal{E}$ is the set difference operator.  $\lceil \cdot  \rceil$ represents the ceiling operator.  The empty set is represented by $\emptyset$.   $\langle{\textbf{u}},\textbf{v}\rangle$ denotes the dot product of vectors $\textbf{u}$ and $\textbf{v}$. $\mbox{blkdiag}(\cdot)$ stands for the block diagonal matrix. The main notations that appear throughout this paper are described in Table \ref{mysymbols}.

\section{System Model}\label{Sys}
\begin{figure}[h!]
	\centering
	\includegraphics[width=0.65\textwidth]{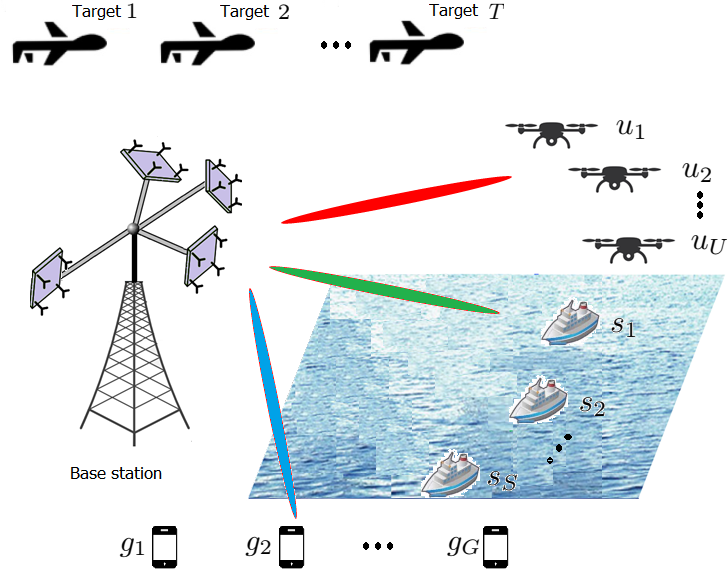}
	\caption{ISAC air–sea-ground  network with movable antenna at the base station.}
	\label{Sys_Mode}
\end{figure}

As shown in Fig. \ref{Sys_Mode}, an 
air–sea-ground  network is considered in which  a base station communicates with a set of  $D$ heterogeneous single-antenna devices and senses a set of $T$ targets $\mathcal{T} = \{\tau_t\}_{t=1}^{T}$.  The heterogeneous devices $\mathcal{D}\triangleq\{\mathcal{U}, \mathcal{S},   \mathcal{G}\}$  consists of  $U$ UAVs $\mathcal{U} = \{u_i\}_{i=1}^{U}$, $S$ sea surface stations $\mathcal{S} = \{s_i\}_{i=1}^{S}$, and  $G$ ground users $\mathcal{G} = \{g_i\}_{i=1}^{G}$ such that $D=U+S+G$. Each device is located at $\bar{\bm{\psi}}_i = [\bar{x}_i, \bar{y}_i, \bar{z}_i]^T$, $\forall i \in \mathcal{D} \cup \mathcal{T}$.
The base station is   equipped with  a set $\mathcal{K}$ of $K$ antenna sub-arrays and  its reference position serves as the origin of the global Cartesian coordinate system (CCS). The total number of antenna elements is $N=\sum_{k=1}^{K}N_k$, with $N_k$ as the number of antenna elements in    the $k$-th sub-array. The sub-arrays
are connected to the base station via
  extendable and rotatable rods, 
and thus, their 3D positions and 3D rotations are adjustable.
 Consequently,
 the  location and   rotation
 of the $k$-th    sub-array, $k \in \mathcal{K}$, can be represented
by six parameters, i.e.,     $\bm{\psi}_k$ for the 3D location and $\bm{\varphi}_k$  for the
3D rotation, that can be defined as 
\begin{equation} 
	\begin{split}
	\bm{\psi}_k = [x_k, y_k, z_k]^T \in \mathcal{C};~	\bm{\varphi}_k = [\alpha_k, \beta_k, \gamma_k]^T,	
	\end{split}
\end{equation}
where  $\mathcal{C}$ is the feasible 3D space around the base station that a sub-array can be placed on; $(x_k, y_k, z_k)$  represents the placement coordinates of the $k$-th    sub-array; and $\alpha_k, \beta_k,$ and $  \gamma_k$ denote the rotation angles around  
the $x$-axis, $y$-axis and $z$-axis, respectively. It is worth mentioning that $\bm{\psi}_k$ is considered as the center point of the $k$-th sub-array as well as the origin of its local  CCS.  Let $ \tilde{\textbf{r}}_{k,n}$ be the position of the $n$-th antenna in the $k$-th sub-array with respect to  the origin of its local CCS; the position of the $n$-th antenna in the $k$-th sub-array with respect to the global CCS  can be expressed as 
\begin{equation}
	\textbf{r}_{k,n}(\bm{\psi}_k,\bm{\varphi}_k) = \bm{\psi}_k + 	\textbf{R}\left(\bm{\varphi}_k\right) \tilde{\textbf{r}}_{k,n}, 
\end{equation}
where  $\textbf{R}\left(\bm{\varphi}_k\right)$ is the rotation matrix  as shown in \eqref{Rotat} \cite{shao2025tutorial}.
\begin{table*}[h]  
	{\footnotesize  
		\begin{equation}\label{Rotat}
			\begin{split}
				\textbf{R}\left(\bm{\varphi}_k\right)\! =\! \begin{blockarray}{ccc}
					\begin{block}{[ccc]}
						\cos(\beta_k) \cos(\gamma_k)& \cos(\beta_k) \sin(\gamma_k) & -\sin(\beta_k) \\
						\sin(\beta_k)\sin(\alpha_k)\cos(\gamma_k) -\cos(\alpha_k)\sin(\gamma_k)&\!\!\!\! \sin(\beta_k)\sin(\alpha_k)\sin(\gamma_k)+\cos(\alpha_k)\cos(\gamma_k) &\!\!\!\! \cos(\beta_k)\sin(\alpha_k)   \\
						\cos(\alpha_k)\sin(\beta_k)\cos(\gamma_k)+ \sin(\alpha_k)\sin(\gamma_k) &\!\!\!\! \cos(\alpha_k)\sin(\beta_k)\sin(\gamma_k)-\sin(\alpha_k)\cos(\gamma_k) &\!\!\!\! \cos(\alpha_k)\cos(\beta_k)   \\
					\end{block}
				\end{blockarray}
			\end{split}
		\end{equation} 
	} 
\end{table*} 

It is worth noting that the outward normal vector of the $k$-th sub-array can be
obtained  as 
${\textbf{n}}_k\left(\bm{\varphi}_k\right)=\textbf{R}\left(\bm{\varphi}_k\right) \tilde{\textbf{n}}_k $,  where  $\tilde{\textbf{n}}_k$ is the normal vector in the local CCS of the $k$-th sub-array. The steering vector of the $k$-th sub-array
is a function of its position $\bm{\psi}_k$   and $\bm{\varphi}_k$ rotation, which can be expressed as

\begin{equation}
	\textbf{a}(\!\bm{\psi}_k,\!\bm{\varphi}_k\!)\!=\!\left[e^{-j {\frac{2\pi}{\lambda}}  \textbf{f}_k\textbf{r}_{k,1}\!(\!\bm{\psi}_k,\!\bm{\varphi}_k\!) },\! \cdots\!, e^{-j {\frac{2\pi}{\lambda}} \textbf{f}_k\textbf{r}_{k,N_k}\!(\!\bm{\psi}_k,\!\bm{\varphi}_k\!) }\right]^T,
\end{equation}
where $\textbf{f}_k$ is   the
pointing vector along the direction of departure $(\theta_k, \phi_k)$, which is defined as $\textbf{f}_k=$ $[\cos(\theta_k)\cos(\phi_k), \cos(\theta_k)\sin(\phi_k), \sin(\theta_k)]$, with  $\theta_k= \arctan(\frac{\dot{y}_k}{\dot{x}_k})$ and $\phi_k = \arctan(\frac{\sqrt{\dot{x}_k^2+\dot{y}_k^2}}{\dot{z}_k})$, where $\dot{x}_k $, $\dot{y}_k$, and $\dot{z}_k$ are the Cartesian components of the  outward normal vector ${\textbf{n}}_k\left(\bm{\varphi}_k\right)$. 

The  
directive   gain of the $k$-th antenna sub-array towards the $i$-th device can be defined as
 \cite{10891142}
\begin{equation}
	\chi_{k,i}= \mathcal{A}-\min\{-(\mathcal{A}_{h_{k,i}}+\mathcal{A}_{v_{k,i}}),\varpi_1\},
\end{equation}
where $\mathcal{A}$ is the maximum gain of the antenna  and $\varpi_1$ represents the front-back ratio.   The horizontal and vertical radiation patterns can be respectively expressed as follows

\begin{equation}
	\begin{split}
&\mathcal{A}_{h_{k,i}} = -\min\{12\left(\frac{\tilde{\phi}_{k}-\tilde{\phi}_{k,i}}{\phi_{3\mbox{\scriptsize dB}}}\right)^2,
 \varpi_1 \},\\
&\mathcal{A}_{v_{k,i}} = -\min\{12\left(\frac{\tilde{\theta}_{k}-\tilde{\theta}_{k,i}}{\theta_{3\mbox{\scriptsize dB}}}\right)^2,
\varpi_2 \}, 
	\end{split}
\end{equation}
where $\phi_{3\mbox{\scriptsize dB}} $  and $\theta_{3\mbox{\scriptsize dB}} $ represent  the horizontal and vertical  $3$-dB beamwidth, respectively,  $\varpi_2$ is the sidelobe level, $(\tilde{\theta}_k, \tilde{\phi}_k)$  represents the direction of departure in the local CCS, and $(\tilde{\theta}_{k,i}, \tilde{\phi}_{k,i})$ is the direction of the $i$-th device/target in the local CCS.

\subsection{Communication Channels} \label{commChannel}

The communication channel between each   antenna sub-array and  a device    depends   on the location
of the device,  the 3D placement position of the  sub-array, and  its 3D rotation along each axis.
Let $\bm{\Psi} \triangleq [\bm{\psi}_1^T, \cdots, \bm{\psi}_K^T]^T$ and $\bm{\Upsilon} \triangleq [\bm{\varphi}_1^T, \cdots, \bm{\varphi}_K^T]^T$; 
the communication   channel  between  the base station and a device $i$   can be modelled  as 

\begin{equation}\label{ComChanel}
	\begin{split}
			\textbf{h}_i(\bm{\Psi},\bm{\Upsilon})=& \left[\Lambda_{1,i}(\bm{\psi}_k)\sqrt{\chi_{1,i}(\bm{\varphi}_1)}\textbf{a}(\bm{\psi}_1,\bm{\varphi}_1)^T,  \cdots,\right.\\
		&\left.  \Lambda_{K,i}(\bm{\psi}_k)\sqrt{\chi_{K,i}(\bm{\varphi}_K)}\textbf{a}(\bm{\psi}_K,\bm{\varphi}_K)^T\right]^T,
	\end{split}
\end{equation}
where    $\Lambda_{k,i}(\bm{\psi}_k)$ is the channel gain between the  $k$-th sub-array and device $i$, which depends on the type of  device.

The channel gain between the  $k$-th sub-array and UAV $i$ $\forall i \in \mathcal{U},$ can be expressed as \cite{8663615}
	\begin{equation}\label{ChannelGainU}
	\Lambda_{k,i} =  \left({\Pr}_{k,i}\left(\mbox{\small LoS}\right)+\left[1-{\Pr}_{k,i}\left(\mbox{\small LoS}\right)\right]\vartheta_0\right)\vartheta_u \delta_{k,i}^{-\ell_u}, 
\end{equation} 
where $\delta_{k,i}=\| \bm{\psi}_k- \bar{\bm{\psi}}_i\|_2 $ is the distance between the $k$-th sub-array and the $i$-th device,  $\vartheta_u$ represents the path loss at  $1$ meter, $\vartheta_0 <1$   is the NLoS attenuation factor,  $\ell_u$ is the path loss exponent,   and  
$ {\Pr}_{k,i}\left(\mbox{\small LoS}\right)$ is the probability of a LoS connection, which is expressed as  
\begin{equation}
	\begin{split}
		{\Pr}_{k,i}\left(\mbox{\small LoS}\right)= \frac{1}{1+ a_1 \exp(-a_2\left[\bar{\theta}_{i,k} - a_1\right])},
	\end{split}
\end{equation}
where 
$ a_1 $ and $ a_2 $ are constants depending on the    type of   environment, while $ \bar{\theta}_{i,k} $ is the elevation angle of the $i$-th UAV   with respect to the $k$-th  sub-array.

The channel gain between the  $k$-th sub-array and surface station $i$ can be expressed as \cite{8528349}
\begin{equation}\label{ChannelGainS}
	\Lambda_{k,i} =  \left[2 \sin \left(\frac{2 \pi h_k h_i}{\lambda \delta_{k,i}}\right)\right]^2\vartheta_s \delta_{k,i}^{-2}, \forall i \in \mathcal{S},
\end{equation} 
where $\vartheta_s$ is the path loss at  $1$ meter,       $ h_i$ represents the height of  the
$i$-th surface station, and $h_k=z_k$ is the height of the $k$-th sub-array.

The  gain of the channel between the  $k$-th sub-array and ground user $i$ can be expressed as 
\begin{equation}\label{ChannelGainG}
	\Lambda_{k,i} =  \left[\sqrt{\frac{F}{F+1}}
	+\sqrt{\frac{1}{F+1}}\tilde{{h}}^{\mbox{\scriptsize NLoS}}_{k,i}\right]\vartheta_g \delta_{k,i}^{-\ell_g}, \forall i \in \mathcal{G},
\end{equation} 
where $\vartheta_g$ is the path loss at  $1$ meter, 
$ \ell_g $ represents the path loss exponent,   $ F $   is the
Rician factor, and $ \tilde{{h}}^{\mbox{\scriptsize NLoS}}_{k,i} \sim \mathcal{N}(0,1)  $ represents the NLoS   scattering component.  The  SINR   at device $i$ can be expressed as

\begin{equation}
		\begin{split}
&	\Gamma_i(\bm{\Psi},\bm{\Upsilon},\mathbf{w})= \\ &\frac{|\mathbf{w}_i^H\textbf{h}_i(\bm{\Psi},\bm{\Upsilon})|^2}{\sum\limits_{\substack{\iota=1\\\iota\neq i}}^{D} |\mathbf{w}_\iota^H\textbf{h}_i(\bm{\Psi},\bm{\Upsilon})|^2\!+
		\sum\limits_{\substack{t=1}}^{T} |\mathbf{\bar{w}}_t^H\textbf{h}_i(\bm{\Psi},\bm{\Upsilon})|^2\!+ \!\sigma^2_i},
			\end{split}
\end{equation} 
where  $\mathbf{w}_i$ is the communication beamformer for the $i$-th device,  $\mathbf{\bar{w}}_t$ is  transmit  beamformer for sensing the $t$-th target,   $\mathbf{w}\triangleq[\mathbf{w}_1^T,\cdots, \mathbf{w}_D^T, \mathbf{\bar{w}}_1^T,\cdots,\mathbf{\bar{w}}_T^T]^T$,  and $\sigma^2_i$ is the power of the additive noise. 
The achievable sum data rate of all devices can be written as 
\begin{equation}\label{Ratee}
	\begin{split}
		&  \mathcal{R}(\bm{\Psi},\bm{\Upsilon},\mathbf{w})= \sum\limits_{\substack{ i=1}}^D \log_2 \left(1 +\Gamma_i(\bm{\Psi},\bm{\Upsilon},\mathbf{w})\right).
	\end{split}
\end{equation}

The sensing
channel of target $\tau_t$   is modelled  as

	
	\begin{equation}\label{ComChanelTarget}
		\begin{split}
			&	\textbf{H}_t(\bm{\Psi},\bm{\Upsilon})=\mbox{blkdiag} \left(\Lambda_{1,t}(\bm{\psi}_k)\sqrt{\chi_{1,t}(\bm{\varphi}_1)}\textbf{a}(\bm{\psi}_1,\bm{\varphi}_1)\textbf{a}(\bm{\psi}_1,\bm{\varphi}_1)^T, \cdots, \right.\\
			&\left.  \Lambda_{K,t}(\bm{\psi}_k)\sqrt{\chi_{K,t}(\bm{\varphi}_1)}\textbf{a}(\bm{\psi}_K,\bm{\varphi}_K)\textbf{a}(\bm{\psi}_K,\bm{\varphi}_K)^T\right)^T,
		\end{split}
	\end{equation}
	where    $\Lambda_{k,t}(\bm{\psi}_k)$ represents the complex amplitude of the target at the $k$-th sub-array, which depends on the path loss and the target's radar cross-section \cite{10158711}.
	The sensing
	channel $\textbf{H}_t(\bm{\Psi},\bm{\Upsilon})$ contains the geometric parameters of
	the target $\tau_t$, such as distance and angle information.
	Let $\mathbf{q}\triangleq[\mathbf{q}_1^T,\cdots, \mathbf{{q}}_T^T]^T$, where $\mathbf{q}_t$ is the designed
	receive beamformer   for sensing the $t$-th target $\tau_t$. The corresponding SINR of   target $\tau_t$ can be written as

	\begin{equation}
		\begin{split}
		&\bar{\Gamma}_t(\bm{\Psi},\bm{\Upsilon},\mathbf{w},\mathbf{q})= \frac{\mathbf{q}_t^H\textbf{H}_t(\bm{\Psi},\bm{\Upsilon})\mathbf{R}\textbf{H}_t^H(\bm{\Psi},\bm{\Upsilon})\mathbf{q}_t}{\mathbf{q}_t^H\left(\sum\limits_{\substack{\varsigma=1\\\varsigma\neq t}}^{T} \textbf{H}_\varsigma(\bm{\Psi},\bm{\Upsilon})\mathbf{R}\textbf{H}_\varsigma^H(\bm{\Psi},\bm{\Upsilon})\!+ \!\sigma^2_r\mathbf{I}_N\right)\mathbf{q}_t},
		\end{split}
	\end{equation} 
	where $\mathbf{R}\triangleq \sum\limits_{\substack{i=1}}^{D}\mathbf{w}_i\mathbf{w}_i^H +\sum\limits_{\substack{t=1}}^{T}\mathbf{\bar{w}}_t\mathbf{\bar{w}}_t^H$  and $\sigma^2_r$ is the power of additive noise.  It has been
	proved in \cite{259642}   that the sensing system   accuracy of estimating the geometric parameters
	of targets  improves as the sensing rate increases, also known as the sensing mutual information
	per unit time.   The     sum sensing rate of all targets can be expressed as 
	\begin{equation}\label{Ratee}
		\begin{split}
			&  \bar{\mathcal{R}}(\bm{\Psi},\bm{\Upsilon},\mathbf{w},\mathbf{q})= \sum\limits_{\substack{ t=1}}^T \log_2 \left(1 +\bar{\Gamma}_t(\bm{\Psi},\bm{\Upsilon},\mathbf{w},,\mathbf{q})\right).
		\end{split}
	\end{equation}

\section{Problem Formulation} \label{ProbFormulation}
The objective is to maximize both the communication sum data rate of all devices and the   sensing sum rate of all targets. Consequently, the objective function is formulated as a weighted sum of the two objectives with $0 \leq\omega\leq 1$ as the relative weight to give preference to improve the communication sum data rate or to improve the sensing sum rate.
 It is worth noting that it is desirable to orient each sub-array outward from the base station's center and to avoid inter-sub-array blockage. Consequently, 
 the optimization problem is formulated as follows:

\begin{subequations}\label{P1}
	\begin{alignat}{2}
		\mbox{\textbf{P1}}~	&\!\max_{\substack{\bm{\Psi},\bm{\Upsilon}\\ \mathbf{w},\mathbf{q}}}&&\!\!\omega\mathcal{R}(\bm{\Psi},\bm{\Upsilon},\mathbf{w}) +(1-\omega)\bar{\mathcal{R}}(\bm{\Psi},\bm{\Upsilon},\mathbf{w},\mathbf{q}), \label{OP1}\\
		&\mbox{s.t.}&& \sum\limits_{\substack{ i=1}}^D\norm{\mathbf{w}_i}^2 +\sum\limits_{\substack{ t=1}}^T\norm{\mathbf{\bar{w}}_t}^2 \leq P,  \label{con2}\\
				&&~~   &  \norm{\mathbf{q}_t}^2 =1, \forall \tau_t \in \mathcal{T}, \label{con33}\\
		&&~~   &   \norm{\bm{\psi}_k-\bm{\psi}_{\bar{k}}}_2 \geq \delta_{\min}, \forall k, \bar{k} \in \mathcal{K},
		\label{con3}\\
		&&~~   &   \langle{\textbf{n}}_k\left(\bm{\varphi}_k\right),\bm{\psi}_k\rangle \geq 0, \forall k \in \mathcal{K},
		\label{con4}\\
		&&~~   &   \langle{\textbf{n}}_k\left(\bm{\varphi}_k\right),\bm{\psi}_{\bar{k}}-\bm{\psi}_k\rangle \leq 0, \forall k, \bar{k} \in \mathcal{K},
		\label{con5}\\
			&&~~   &   \bm{\psi}_k \in \mathcal{C}, \forall k \in \mathcal{K}.
		\label{con6}
	\end{alignat}
\end{subequations}
 Constraint \eqref{con2} represents the power budget at the base station.
  Constraint \eqref{con33}
 normalizes the receive beamformers. 
  Constraint \eqref{con3} guarantees a minimum distance separation among sub-arrays to avoid overlapping and coupling.  
Constraint \eqref{con4} guarantees that each sub-array is pointing outward from the base station's center, where $\langle{\textbf{u}},\textbf{v}\rangle$ is the dot product of vectors $\textbf{u}$ and $\textbf{v}$.\footnote{The dot product of two vectors $\textbf{u}$ and $\textbf{v}$ can be expressed as $\langle{\textbf{u}},\textbf{v}\rangle=\norm{\textbf{u}}\norm{\textbf{u}} \cos(\theta_{\textbf{u},\textbf{v}})$, and   is a positive value if the angle between the vectors $\theta_{\textbf{u},\textbf{v}}$ is an acute angle; on the other hand,   it is a negative value if the angle is obtuse.} Constraint \eqref{con5} is introduced to avoid the mutual blockage among the sub-arrays. Constraint \eqref{con6} guarantees that   each
sub-array is located in the given base station’s 3D site space $\mathcal{C}$.

\section{Solution Approach}\label{Sol2}
This section discusses the     designed solution to solve the formulated optimization problem. The receive and transmit beamforming   are designed   using the generalized eigenvector and successive convex approximation approaches, respectively.    The rotations of the sub-arrays $\bm{\Upsilon}$ are obtained using a $K$-means clustering approach, while a PSO technique is developed to find the locations of the sub-arrays $\bm{\Psi}$.

	\subsection{Optimizing the Receive Beamforming}\label{Beam_R}
	
	For given $\bm{\Psi},\bm{\Upsilon}$, and $ \mathbf{w}$,  the receive beamformer $\mathbf{q}_t$ only affects the sensing rate of the target $\tau_t$, which is an increasing function with the corresponding SINR. Consequently, the optimum   receive beamformer $\mathbf{q}^*_t$ is the solution of the following optimization problem

	\begin{subequations}\label{P1_q}
		\begin{alignat}{2}
			\mbox{\textbf{P1.Q}}~	&\!\max_{\substack{\mathbf{q}_t}}&&\frac{\mathbf{q}_t^H\textbf{H}_t\mathbf{R}\textbf{H}_t^H\mathbf{q}_t}{\mathbf{q}_t^H\left(\sum\limits_{\substack{\varsigma=1\\\varsigma\neq t}}^{T} \textbf{H}_\varsigma\mathbf{R}\textbf{H}_\varsigma^H\!+ \!\sigma^2_r\mathbf{I}_N\right)\mathbf{q}_t}, \label{OP1q}\\
			&\mbox{s.t.}&&   \norm{\mathbf{q}_t}^2 =1.
		\end{alignat}
	\end{subequations}
	It is worth noting that the objective  in \eqref{OP1q} is a generalized Rayleigh quotient which can be written as $\frac{\mathbf{q}_t^H\mathbf{A}_t\mathbf{q}_t}{\mathbf{q}_t^H\mathbf{B}_t\mathbf{q}_t} $, where $\mathbf{A}_t=\textbf{H}_t\mathbf{R}\textbf{H}_t^H$ and $\mathbf{B}_t=\sum\limits_{\substack{\varsigma=1\\\varsigma\neq t}}^{T} \textbf{H}_\varsigma\mathbf{R}\textbf{H}_\varsigma^H\!+ \!\sigma^2_r\mathbf{I}_N $. After simple manipulations, it can be shown that the global
	maximizer of \eqref{OP1q} is the  largest generalized eigenvector of $(\mathbf{A}_t,\mathbf{B}_t)$. Finally, keeping in mind that $\eqref{OP1q}$ is a scaling
	invariant, the optimum receive beamformer $\mathbf{q}^*_t$ is the normalized largest generalized eigenvector of $(\mathbf{A}_t,\mathbf{B}_t)$.

	\subsection{Optimizing the Transmit Beamforming}\label{Beam}
	
	For given $\bm{\Psi},\bm{\Upsilon}$, and $ \mathbf{q}$, the transmit beamformer can be designed by solving the following optimization
	problem

	\begin{subequations}\label{P1W}
		\begin{alignat}{2}
			\mbox{\textbf{P1.W}}~	&\!\max_{\substack{\mathbf{w}}}&&~\omega\mathcal{R}(\mathbf{w}) +(1-\omega)\bar{\mathcal{R}}(\mathbf{w}), \label{OP1}\\
			&\mbox{s.t.}&& \sum\limits_{\substack{ i=1}}^D\norm{\mathbf{w}_i}^2 +\sum\limits_{\substack{ t=1}}^T\norm{\mathbf{\bar{w}}_t}^2 \leq P.  \label{con2W}
		\end{alignat}
	\end{subequations}
It can be noticed that $1+\Gamma_i(\mathbf{w})$ can be written as $\frac{\mathbf{w}^H\mathbf{C}_i\mathbf{w}}{\mathbf{w}^H\mathbf{D}_i\mathbf{w}}$, where 
\begin{equation}
	\begin{split}
		&\mathbf{C}_i=\mbox{blkdiag}(\mathbf{h}_i\mathbf{h}^H_i, \cdots, \mathbf{h}_i\mathbf{h}^H_i)+ \sigma_i^2\mathbf{I}_{N(D+T)},\\
		&	\mathbf{D}_i=\mathbf{C}_i-\mbox{blkdiag}(\mathbf{0}, \cdots, \mathbf{h}_i\mathbf{h}^H_i, \cdots, \mathbf{0}).
	\end{split}
\end{equation}
Furthermore, $1+\bar{\Gamma}_t(\mathbf{w})$ can be written as $\frac{\mathbf{w}^H\mathbf{\bar{C}}_t\mathbf{w}}{\mathbf{w}^H\mathbf{\bar{D}}_t\mathbf{w}}$, where 
\begin{equation}
	\begin{split}
		&\mathbf{\bar{C}}_t=\mbox{blkdiag}(\mathbf{H}_t^H\mathbf{q}_t\mathbf{q}_t^H\mathbf{H}_t, \cdots, \mathbf{H}_t^H\mathbf{q}_t\mathbf{q}_t^H\mathbf{H}_t)\\
		&~~~~~~ + \sigma_r^2\norm{\mathbf{q}_t}^2\mathbf{I}_{N(D+T)},\\
		&	\mathbf{\bar{D}}_t=\mathbf{\bar{C}}_t-\mbox{blkdiag}(\mathbf{0}, \cdots, \mathbf{H}_t^H\mathbf{q}_t\mathbf{q}_t^H\mathbf{H}_t, \cdots, \mathbf{0}).
	\end{split}
\end{equation}

	By introducing  the
	auxiliary variables   $\bm{\lambda}\triangleq[\lambda_1, \cdots, \lambda_D, \bar{\lambda}_1, \cdots, \bar{\lambda}_T]$ and the  exponential auxiliary variables  $e^{c_i}$, $e^{d_i}$, $e^{\bar{c}_t}$, and $e^{\bar{d}_t}$, the optimization problem \eqref{P1W} can be rewritten as

\begin{subequations}\label{P1_approx}
	\begin{alignat}{2}\nonumber
		\mbox{\textbf{P2.W
		}}~	&\!\!\max_{\substack{\mathbf{w},\bm{\lambda}\\c_i,d_i,\bar{c}_t,\bar{d}_t}} \!&&  \omega\sum_{i=1}^{D} \lambda_i + (1-\omega)\sum_{t=1}^{T} \bar{\lambda}_t \\
		&\mbox{s.t.}&&  c_i-d_i \geq \lambda_i \log2,\forall i \in \mathcal{D},  \label{con1-21}\\
		&&~~   & \bar{c}_t-\bar{d}_t \geq \bar{\lambda}_t \log2, \forall t \in \mathcal{T},  \label{con1-222}\\
		&&~~   &
		e^{c_i} \leq	\mathbf{w}^H\mathbf{C}_i\mathbf{w}, \\ 
		&&~~   & e^{d_i} \geq	\mathbf{w}^H\mathbf{D}_i\mathbf{w}, \label{Conn1} \\ 
		&&~~   & e^{\bar{c}_t} \leq	\mathbf{w}^H\mathbf{\bar{C}}_t\mathbf{w}, \\ 
		&&~~   & e^{\bar{d}_t} \geq	\mathbf{w}^H\mathbf{\bar{D}}_t\mathbf{w}, \label{Conn2}\\ 
		&&~~   &  \eqref{con2}.
	\end{alignat}
\end{subequations}

It can be noticed that  \eqref{Conn1} and \eqref{Conn2} are non-convex.  By utilizing the first-order Taylor expansion, \eqref{Conn1} and \eqref{Conn2} can be respectively   approximated as 
\begin{equation}\label{Linearlization1} 
	\begin{split}
		&e^{\tilde{d}_i{(k)}}(d_i-\tilde{d}_i{(k)} +1) \geq	\mathbf{w}^H\mathbf{D}_i\mathbf{w},\\
		& e^{\tilde{\bar{d}}_t{(k)}} (\bar{d}_t-\tilde{\bar{d}}_t{(k)} +1) \geq	\mathbf{w}^H\mathbf{\bar{D}}_t\mathbf{w},
	\end{split}
\end{equation}
where $\tilde{d}_i{(k)}$ and  $\tilde{\bar{d}}_t{(k)}$   represent the points around which the linearizations are made at the $k$-th iteration; these parameters are updated iteratively  as shown in Algorithm \ref{AlgOpti}, which illustrates the SCA approach to solve \eqref{P1_approx}.

\begin{algorithm}
	\caption{Successive convex approximation algorithm for solving problem \eqref{P1_approx}.}\label{AlgOpti}
	\begin{algorithmic}[1] 
		\State  \textbf{Input:} $\mathbf{C}_i$, $\mathbf{D}_i$,  $\forall i \in \mathcal{D}$ $\mathbf{\bar{C}}_t$, $\mathbf{\bar{D}}_t$,  $\forall t \in \mathcal{T}$;
		\State  Initialize   $\tilde{d}_i{(0)}$, and  $\tilde{\bar{d}}_t{(0)}$; Set $j=1$;
		\State $ ~~ $\textbf{While}  the   improvement in the objective  value is above   a predefined threshold $\epsilon$  \textbf{and} $j \leq J$ \textbf{do}
		\State   Obtain $\mathbf{w}^*,\bm{\lambda}^*,c_i^*,d_i^*,\bar{c}_t^*,$ and $\bar{d}_t^*$ by solving \eqref{P1_approx} with \eqref{Linearlization1};
		\State $ ~~~ $   $\tilde{d}_i{(k)}\leftarrow d_i^*$;
		\State $ ~~~ $   $\tilde{\bar{d}}_t{(k)}\leftarrow\bar{d}_t^*$;	$j=j+1$; 
		\State $ ~~ $\textbf{End While} 
		\State  \textbf{Return}  $\mathbf{w}^*$.
		\normalsize
	\end{algorithmic}
\end{algorithm}

\begin{algorithm} 
	\caption{ $K$-means clustering for optimized  sub-arrays' rotation.}\label{KmeansClustering}
	\begin{algorithmic}[1]
		\State \textbf{Input:} Number of sub-arrays $K$, $\omega$, $\bar{\bm{\psi}}_i $ $\forall i \in \mathcal{D}$, and maximum number of iteration $\bar{J}$;  
		\State  Select $K=\bar{D}+\bar{T}$ devices/targets, where $\bar{D}=\lceil \omega K \rceil$ devices and $\bar{T} = K - \bar{D}$ targets;  Set the locations of the selected devices/targets as seeds  for the cluster centroides $c_1(1), c_2(1), \cdots, c_K(1)$; Set $j=1$;
    	\State \textbf{If} $\omega =0.5$
		\State $ ~ $  $\bar{\omega}_i = 1$, $\forall i \in \mathcal{D}\cup \mathcal{T}$;
		\State \textbf{Else If} $\omega >0.5$
		\State $ ~ $  $\bar{\omega}_i = 1$, $\forall i \in \mathcal{D}$; $\bar{\omega}_i = 1-\omega$, $\forall i \in \mathcal{T}$;
		\State \textbf{Else If} $\omega <0.5$
		\State $ ~ $  $\bar{\omega}_i = 1-\omega$, $\forall i \in \mathcal{D}$; $\bar{\omega}_i = 1$, $\forall i \in \mathcal{T}$;
				\State \textbf{End If}
		\State $ ~ $\textbf{While} $\norm{c_k(j)-c_k(j-1)} \geq \epsilon ~ \forall k=1, \cdots, K$   \textbf{and}  $j \leq \bar{J}$ \textbf{do}
		\State $~~$  $ \bm{\mu} \leftarrow [\bm{0}]_{K\times (D+T)}$;
		\State $ ~~~ $\textbf{For}  $i =1$ \textbf{to} $D+T$  \textbf{do}
		\State $~~~~$ $ \kappa=\arg \min\limits_{1\leq k\leq K} \{\norm{c_k(j)-\bar{\bm{\psi}}_i}\}$;
		\State $~~~~$  $\mu_{\kappa,i}=1$;
		\State $ ~~~ $\textbf{End For}
		\State $~~~$ Set $j=j+1$;
		\State $ ~~~ $ \textbf{For}  $k =1$ \textbf{to} $K$  \textbf{do}
		\State $ ~~~ $	  $c_k(j) = \frac{\sum\limits_{i=1}^{D+T}{\bar{\omega}_i}\mu_{k,i} \bar{\bm{\psi}}_i}{\sum\limits_{i=1}^{D+T}{\bar{\omega}_i}\mu_{k,i}}$;
		\State $ ~~~ $ \textbf{End For}
		\State $ ~~~ $\textbf{End While}
	    \State $~$	$\bm{\eta}_k\leftarrow c_k(j)$,  $\forall k=1, \cdots, K$;
		\State $~$ Obtain the outward normal vectors  of the  sub-arrays in  the global CCS as ${\textbf{n}}_k\triangleq[\dot{x}_k, \dot{y}_k, \dot{z}_k]=\frac{\bm{\eta}_k}{\norm{\bm{\eta}_k}}$,  $\forall k=1, \cdots, K$; 
	    \State $~$ Obtain the rotation angles\footref{refnote} $\alpha_k=\tan^{-1}\left(\frac{\dot{y}_k}{\dot{z}_k}\right)$ and  $\beta_k=\tan^{-1}\left(\frac{\dot{x}_k}{\dot{z}_k}\right)$  $\forall k=1, \cdots, K$;
		\State \textbf{Return} ${\textbf{n}}_k$   $\bm{\eta}_k$, and $\bm{\varphi}_k = [\alpha_k, \beta_k, 0]^T$, $\forall k=1, \cdots, K$.
		\normalsize
	\end{algorithmic}
\end{algorithm}

\subsection{Optimizing the Rotations of the Sub-arrays }

 To optimize the sub-arrays rotation, it is worth mentioning that the antenna elements are typically configured as a uniform planar array (UPA) in which the main lobe is outwards of one side of the UPA. 
 If a sub-array is placed in the $x$-$y$ plane,  the effect of rotating the sub-array with respect to the $z$-axis is negligible. The original orientation of the unit vector is considered along the positive $z$-axis, i.e., $\textbf{n}=[0, 0, 1]$. Keeping in mind that the reference position of the base station is at the origin of the global CCS, rotating the   unit vector around  the $x$-axis to be $\textbf{n}_x=[0, \dot{y}, \dot{z}]$ can be obtained by setting the rotation angle around  the $x$-axis as\footnote{It is worth noting that   a quadrant adjustment should be applied by adding $\pi$ to the angle if $\dot{z} <0$.\label{refnote}}  $\alpha=\tan^{-1}\left(\frac{\dot{y}}{\dot{z}}\right)$. Similarly, rotating the   unit vector around  the $y$-axis to be $\textbf{n}_y=[\dot{x}, 0, \dot{z}]$ can be obtained by setting the rotation angle around  the $y$-axis as\footref{refnote}  $\beta=\tan^{-1}\left(\frac{\dot{x}}{\dot{z}}\right)$.
 {Algorithm} \ref{KmeansClustering} shows the main steps to find the outward normal vector  of each sub-array in  the global CCS.
The devices and targets are clustered into $K$ clusters, in which  the  centroides are initialized to be the locations of a randomly selected devices. The algorithm iterates and adjusts the centroides until there is no remarkable change in the centroids   or a maximum number of iterations is reached. Assuming the base station is placed at the origin of the global CCS, the outward normal vectors of the sub-arrays in the  global CCS are the unit vectors pointing towards the centroids. 

\begin{algorithm} 
	\caption{Movable antenna for air–sea-ground networks.}\label{euclid1}
	\begin{algorithmic}[1]
		 	\small
		\State \textbf{Input:} The locations of the devices $\bar{\bm{\psi}}_i$,   $\forall i \in \mathcal{D}$,  base station’s 3D site space $\mathcal{C}$, minimum inter sub-array distance $\delta_{\min}$,  number of practicals $\Omega$, step size $\varepsilon$; maximum number of iterations $\bar{\bar{J}}$;
		\State Obtain the   the   outward normal vectors in the global CCS ${\textbf{n}}_k$, the centroides $\bm{\eta}_k$, and the rotation angles $\bm{\varphi}_k$, $\forall k=1, \cdots, K$ using \textbf{Algorithm} \ref{KmeansClustering};
		\State  Initialize the locations of the particles set $\bm{\varOmega}\triangleq\hat{\bm{\psi}}_\kappa \in \mathcal{C}$, $\forall \kappa=1, 2, \cdots, \Omega$; Set  $\bm{\Psi}^* \leftarrow \emptyset$; ${R}^*=0$;
		\State $ ~ $\textbf{While} $j$ $\leq$  $\bar{\bar{J}}$   \textbf{do}
		\State $~~$ Calculate the distance between each particle and centroid $\delta_{\kappa,k}=\norm{\hat{\bm{\psi}}_\kappa-\bm{\eta}_k}_2$, $\forall 1 \leq \kappa\leq \Omega, 1 \leq k \leq K$;
		\State $~~$ $\bm{\varPsi}^* \leftarrow  \emptyset$;  $\bm{\mathcal{V}} \leftarrow  \emptyset$; $k=1$;
		\State $ ~~~ $\textbf{While}  $k \leq K$  \textbf{do}
		\State $~~~~$ $ k^*=\arg \min\limits_{{\forall \kappa \in \bm{\varOmega}\setminus\bm{\mathcal{V}}}} \{\delta_{\kappa,k}\}$;
		\State $~~~~$\textbf{If}  $ \norm{\hat{\bm{\psi}}_{k^*}-\hat{\bm{\psi}}_{\bar{\kappa}}}_2 \geq d_{\min}$ \textbf{and}  
		$\langle{\textbf{n}}_k,\hat{\bm{\psi}}_{k^*}\rangle \geq 0$ \textbf{and} $~~~~~~~$ $~~~~~~~\langle{\textbf{n}}_k,\hat{\bm{\psi}}_{\bar{\kappa}}-\hat{\bm{\psi}}_{k^*}\rangle \leq 0 $,		
		$ \forall \bar{\kappa} \in \bm{\varPsi}^*$    \textbf{do}
		\State $~~~~~$  $\bm{\varPsi}^* \leftarrow \bm{\varPsi}^* \cup \hat{\bm{\psi}}_{k^*}$;
		\State $~~~~~$  Obtain the outward normal vector in the local CCS $\tilde{\textbf{n}}_k\triangleq[\dot{\tilde{x}}_k, \dot{\tilde{y}}_k, \dot{\tilde{z}}_k]=\textbf{R}^{-1}\left(\bm{\varphi}_k\right) \frac{\bm{\eta}_k-\hat{\bm{\psi}}_{k^*}}{\norm{\bm{\eta}_k-\hat{\bm{\psi}}_{k^*}}}$; 
		\State $~~~~~$  Obtain the  vector  $\tilde{\textbf{n}}_{k,i}\triangleq[\dot{\tilde{x}}_{k,i}, \dot{\tilde{y}}_{k,i}, \dot{\tilde{z}}_{k,i}]=\textbf{R}^{-1}\left(\bm{\varphi}_k\right) \frac{\bm{\bar{\psi}}_i-\hat{\bm{\psi}}_{k^*}}{\norm{\bm{\bar{\psi}}_i-\hat{\bm{\psi}}_{k^*}}}$;
		\State $~~~~~$  Obtain the direction of departure in the local CCS as $\tilde{\theta}_k= \arctan(\frac{\dot{\tilde{y}}_k}{\dot{\tilde{x}}_k})$ and $\tilde{\phi}_k = \arctan(\frac{\sqrt{\dot{\tilde{x}}_k^2+\dot{\tilde{y}}_k^2}}{\dot{\tilde{z}}_k})$;
		\State $~~~~~$  Obtain the direction of $i$-th device/target in the local CCS as $\tilde{\theta}_{k,i}= \arctan(\frac{\dot{\tilde{y}}_{k,i}}{\dot{\tilde{x}}_{k,i}})$ and $\tilde{\phi}_{k,i} = \arctan(\frac{\sqrt{\dot{\tilde{x}}_{k,i}^2+\dot{\tilde{y}}_{k,i}^2}}{\dot{\tilde{z}}_{k,i}})$;
		\State $~~~~$ $\bm{\mathcal{V}} \leftarrow  \emptyset$; $k=k+1$;
		\State $~~~~$\textbf{Else}
		\State $~~~~~$ $\bm{\mathcal{V}} \leftarrow \bm{\mathcal{V}} \cup  \hat{\bm{\psi}}_{k^*}$;
		\State $~~~~$\textbf{End If}
		\State $ ~~~ $\textbf{End While}
		\State $ ~~ $\textbf{While}  the   increase in the objective function value is greater than   a predefined threshold $\epsilon$  \textbf{and} $l \leq l^{\mbox{\scriptsize max}}$ \textbf{do}
		\State $~$ Obtain the transmit beamforming $\mathbf{w}$ using Algorithm \ref{AlgOpti};
		\State $~$ Obtain the receive beamforming as the normalized largest generalized eigenvector of $(\mathbf{A}_t,\mathbf{B}_t)$, $\forall\tau_t\in\mathcal{T}$;
		\State $~$ Calculate the corresponding objective value ${R}=\omega\mathcal{R}(\bm{\Psi},\bm{\Upsilon},\mathbf{w}) +(1-\omega)\bar{\mathcal{R}}(\bm{\varPsi}^*,\bm{\Upsilon},\mathbf{w},\mathbf{q})$; 
		\State $~~$ $l=l+1$;
				\State $ ~~~ $\textbf{End While}
		\State $~~$  \textbf{If}  ${R} > {R}^* $
		\State $~~~~$  ${R}^* = {R}$; $\bm{\Psi}^* \leftarrow \bm{\varPsi}^*$;
		\State $~~$ \textbf{End If}
		\State $~~$ Update the locations of the particles $\tilde{\bm{\psi}}_\kappa \leftarrow \varepsilon (\tilde{\bm{\psi}}_\kappa + \tilde{\bm{\psi}}_{\kappa^*})$, where $\tilde{\bm{\psi}}_{\kappa^*}$ is the nearest point to the particle $\tilde{\bm{\psi}}_\kappa$ in $\bm{\varPsi}^*$;
		\State $~~$  $j=j+1$;
		\State $ ~ $\textbf{End While}
		\State \textbf{Return} ${R}^*$.
		\normalsize
	\end{algorithmic}  
\end{algorithm}

\subsection{Optimizing the Positions of the Sub-arrays }
The developed PSO iterates to  improve the candidate positions of the sub-arrays. It solves the problem by having a population of candidate positions  (particles) and moving these particles around in the  given base station’s 3D site space $\mathcal{C}$ to select the best locations. The movement of
each particle is influenced by its  current  position and is   guided toward the location of the nearest best known particle in the particles set.  {Algorithm} \ref{euclid1} illustrates the overall solution approach, which starts by obtaining the intended direction of each sub-array   using the $K$-means clustering (line $2 $). The PSO (lines $3\sim14$) searches for the best feasible locations for the sub-arrays and rotates them towards the intended direction.  The corresponding transmit and receive beamformers are obtained, and the corresponding objective value is calculated (lines $15\sim26$). 

\begin{table}[h]
	\caption{Numerical Values of Parameters.}
	\begin{center}
		\begin{adjustbox}{width=.65\textwidth}
			\begin{tabular}{|c|c||c|c||c|c|}
				\hline
				\textbf{Parameter}&\textbf{Value} & \textbf{Parameter}&\textbf{Value}   & \textbf{Parameter}&\textbf{Value}    \\ \hline
				$ P$ 	&   $ 25 $ W   &  $\varpi_1,\varpi_2 $ 	& $ 20 $ dBi  &    $\theta_{3\mbox{\scriptsize dB}},\phi_{3\mbox{\scriptsize dB}}$    &  $65^{\circ}$  \\ \hline  
				$ \vartheta_u,\vartheta_s,\vartheta_g$ 	&  $-30 $ dB  &  $ \ell_u,\ell_g $	&  $ 2.2 $    &  $a_1,a_2$      &   $10, 0.6$  \\  \hline  
				$ \sigma^2_i$ 	&   $ -110\! $   dBm  &  $F $ 	& $10 $   &     $\epsilon$   &  $0.1$  \\ \hline  
				$ J$ 	&   $ 10$       &  $\varepsilon $	& $0.5$   &    $\mathcal{C}$    &  $27$ m$^3$ \\ \hline  
				$ \bar{J}$ 	&   $ 10$       &  $\bar{\bar{\bar{J}}} $	& $10$   &    $\Omega$   &  $100$  \\ \hline  
				$\omega$ 	&   $ 0.5$       &  $K $	& $8$   &    $N_k$    &  $8$  \\ \hline  	    	   							             	    	   				 	 	 	 	          	    	   							             	    	   				 	 	 	 	          
			\end{tabular}
		\end{adjustbox}
	\end{center}
	\label{TableResults}
\end{table}  

\subsection{Computational Complexity}
Selecting the $K$  cluster seeds in Algorithm \ref{KmeansClustering} requires $\mathcal{O}(K)$ operations. Each iteration of Algorithm \ref{KmeansClustering} involves selecting the nearest cluster for each device (lines $12\sim15$), which requires $\mathcal{O}((D+T)K)$ operations. Updating the clusters' centroids  (lines $17\sim19$)  requires $\mathcal{O}((D+T)K)$ operations.
Consequently, obtaining the intended direction of the sub-arrays using Algorithm \ref{KmeansClustering} requires  $\mathcal{O}(\bar{J}(D+T)K)$  operations. 
The PSO approach  (lines $5\sim20$ of Algorithm \ref{euclid1}) involves calculating the distance between each particle and  centroid (line $6$), which requires  $\mathcal{O}(K\Omega)$ operations.  Selecting the nearest particle for each  centroid that satisfies the conditions in line $10$    requires $\mathcal{O}(K\Omega^2)$ operations.
Designing the receive beamforming requires $\mathcal{O}(TN^3)$ operations, and Algorithm \ref{AlgOpti} requires $\mathcal{O}(JN^{3.5}(D+T)^{3.5})$ operations to find  the transmit beamforming. 
Consequently, obtaining the beamforming (lines $22\sim23$ of Algorithm \ref{euclid1}) requires $\mathcal{O}(l^{\mbox{\scriptsize max}}JN^{3.5}(D+T)^{3.5})$ operations.
Keeping in mind that $N\geq K$, the overall computation complexity of solving problem \textbf{P1} can be expressed as $\mathcal{O}(\bar{J}(D+T)K+ \bar{\bar{J}}( K\Omega^2+l^{\mbox{\scriptsize max}}JN^{3.5}(D+T)^{3.5})$.

\section{Simulation Results}\label{SimSec}
This section      illustrates the performance of the proposed movable antenna-based framework and the developed solution approach. The   numerical values of the  parameters are presented in Table \ref{TableResults}; these values are employed unless
otherwise stated. 
Two  different cases     are considered as follows:
\begin{itemize} 
	\item \textit{The proposed framework},  which is the performance of the solution approach in {Algorithm} \ref{euclid1}. 
	\item \textit{Stationary antenna array},  which represents the performance of the conventional scenario such that  the placement and   rotation
	of all sub-array antennas are fixed and the transmit and receive beamformers are obtained using {Algorithm} \ref{AlgOpti}   and as described in Section \ref{Beam_R}, respectively.
	
\end{itemize}

\begin{figure}[h!] 
	\centering
	\includegraphics[width=0.65\textwidth]{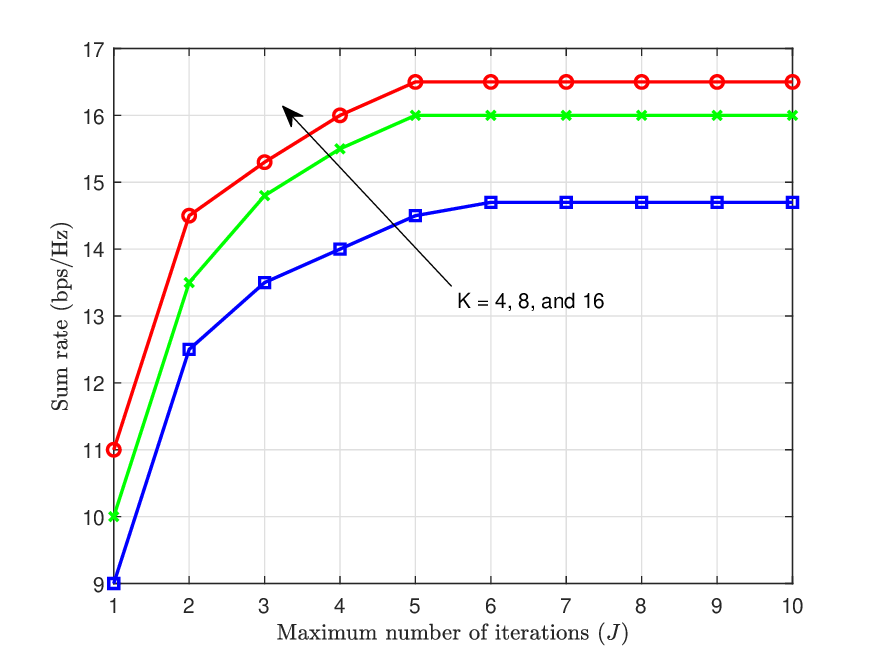}
	\caption{Convergence performance of Algorithm \ref{euclid1}, with $U=6$ UAVs, $S=6$ sea surface stations, $D=6$ ground users, $T=3$ targets,  $N=64$, and $N_k=N/K$.}
	\label{figIteration}
\end{figure}

\begin{figure}[h!] 
	\centering
	\includegraphics[width=0.65\textwidth]{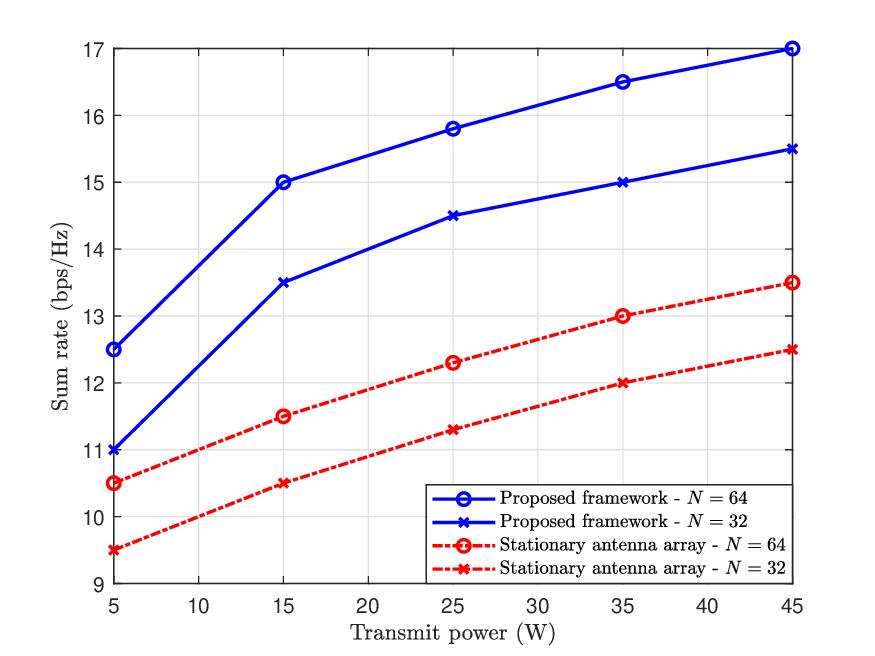}
	\caption{Sum rate versus transmit power	 with $U=6$ UAVs, $S=6$ sea surface stations, $D=6$ ground users, $T=3$ targets,  and $N_k=N/K$.}
	\label{figP}
\end{figure}

Figure \ref{figIteration} illustrates the performance of the developed solution approach versus the maximum number of iterations in Algorithm \ref{euclid1}. This figure also shows the performance of the  developed movable antenna framework with different number of sub-arrays $K$. It can be noticed that the algorithm demonstrates good convergence, and the objective value stabilizes within a few iterations. It can also be noticed that the objective value improves as the number of sub-arrays increases, even though the total number of antenna elements remains fixed. This result is consistent with the observation that increasing the number of sub-arrays adds more degrees of freedom, thereby further improving performance.

Figure \ref{figP} illustrates the objective value versus the  base station's maximum transmit  power. It is noticeable that the objective improves with the transmit power and the developed movable antenna framework provide remarkable improvement in the  overall system performance. It can be also noticed that  adding more    antenna elements improves the performance of both the developed movable antenna framework and the conventional stationary antenna array.

\begin{figure}[h!] 
	\centering
	\includegraphics[width=0.65\textwidth]{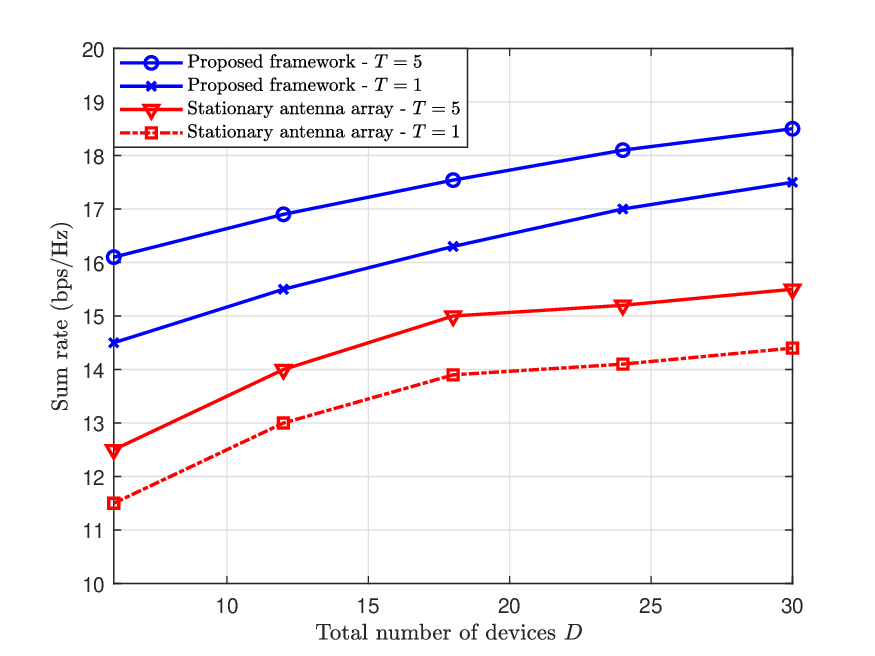}
	\caption{Sum rate versus 	the total number of devices $ D $, with $U=S=G=D/3$,  $N=64$, and $N_k=8$.}
	\label{figD}
\end{figure}

\begin{figure}[h!] 
	\centering
	\includegraphics[width=0.65\textwidth]{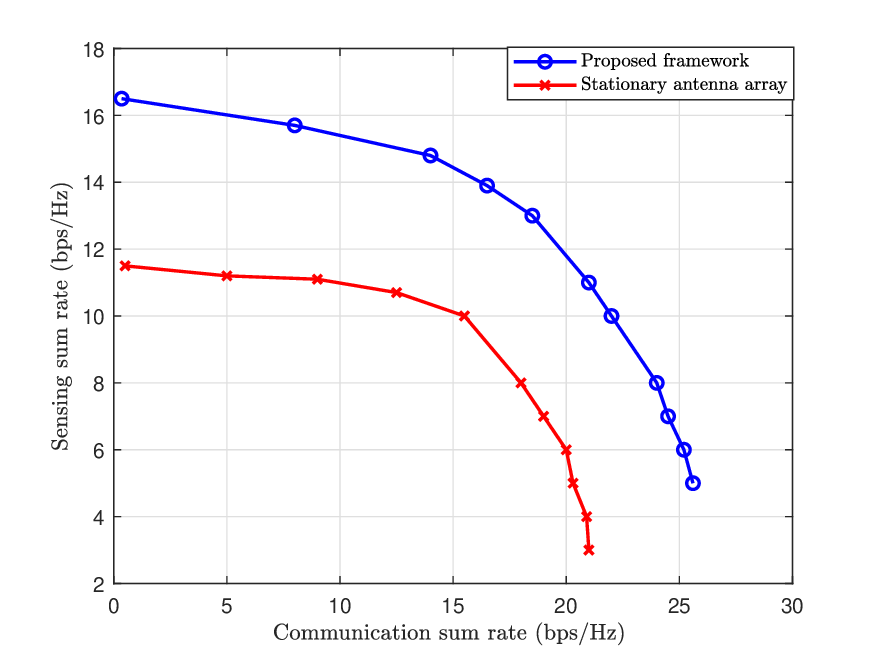}
	\caption{Trade-off between sensing  sum rate and communication sum rate.}
	\label{figTrad}
\end{figure}

Figure  \ref{figD} shows 
the  performance of the proposed movable antenna-based framework and the conventional stationary antenna array  versus a range of number of devices,  with a range   of number of  targets $T$. It can be noticed that the  movable antenna-based framework provides better performance; additionally, as the more devices and targets are involved in the system, the performance   enhances. This result is consistent with the observation    that the additional devices/targets  
leverage a   spatial diversity  and yield better   communication/sensing channel between the base station and the devices/targets.

To study the trade-off between targets' sensing and
data communication, Fig. \ref{figTrad} illustrates the sensing sum rate versus the communication sum rate such that each curve is obtained by changing the value of  the relative weight $0 \leq\omega\leq 1$. It can be seen that the developed movable antenna framework maintains
better overall performance and trade-off between the two objectives,  which gives the decision-maker a
set of satisfactory trade-off solutions for maximizing the  sensing sum rate and the communication sum rate.

\section{Conclusion}\label{CO}
This paper  introduced a novel movable antenna-based framework to enable ISAC in air–sea-ground  networks by optimizing  the transmit and receive beamforming, as well as the location and   orientation of antenna sub-array of a movable antenna   scenario. 
A multi-objective optimization problem was formulated with the objective of maximizing both the sensing sum rate  of a set of targets and the communication data sum rate  of a set of  air–sea-ground devices.
Practical constraints were considered on the the sub-arrays placement and orientation. 
A   solution approach was developed based on $K$-means clustering, PSO,    generalized eigenvector, and successive convex approximation approaches.   Results showed that the developed  movable antenna-based framework   remarkably improves both the sensing sum rate and the communication sum rate when compared with the conventional stationary antenna array.

\balance
\bibliographystyle{IEEEtran}
\bibliography{IEEEabrv,Refrences-library}

\end{document}